\documentclass[letter, twocolumn]{aa}
\usepackage[varg]{txfonts}
\usepackage{graphicx}
\usepackage{graphics}
\usepackage{comment}
\graphicspath{{./figures/}}
\usepackage{natbib}
\usepackage{subfig}
\usepackage{hyperref}
\usepackage{soul}
\usepackage[noabbrev]{cleveref}

\usepackage{xcolor}

\newcommand{\orcid}[1]{\href{https://orcid.org/#1}{\includegraphics[width=10pt]{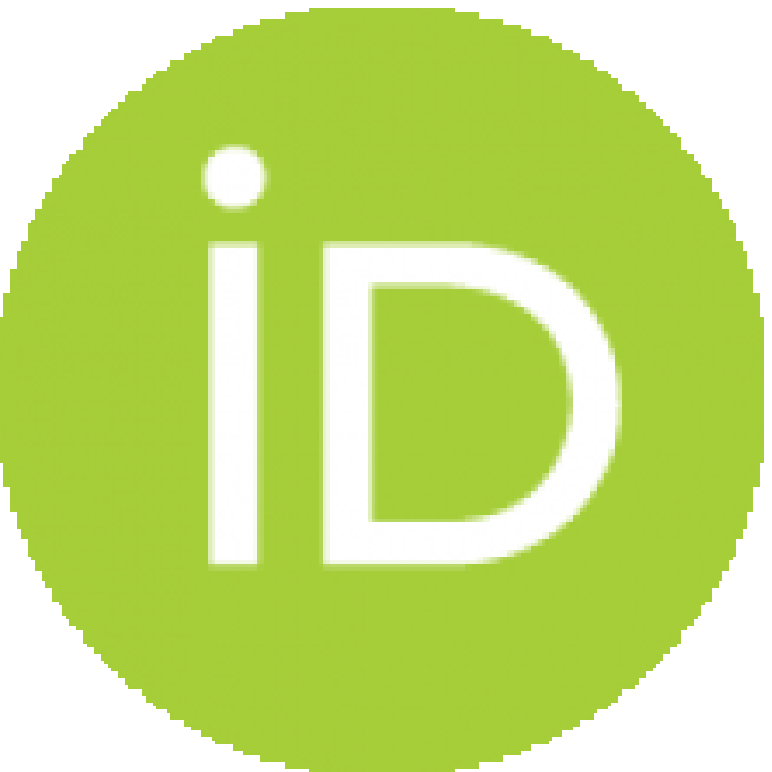}}}

\let\oldhat\hat
\renewcommand{\vec}[1]{\boldsymbol{#1}} 
\renewcommand{\hat}[1]{\oldhat{\boldsymbol{#1}}}
\newcommand{\Msun}{M_{\odot}}
\newcommand{\etaOD}{\eta_{\rm OD,0}}
\newcommand{\etaODt}{\tilde{\eta}_{\rm OD,0}}
\newcommand{\etaAD}{\eta_{\rm AD,0}}

\newcommand{\tauni}{\tau_{ni,0}}
\newcommand{\taunit}{\tilde{\tau}_{ni,0}}
\newcommand{\beq}{\begin{equation}}
\newcommand{\eeq}{\end{equation}}

\begin{document}

\title{Variation of the Core Lifetime and Fragmentation Scale in Molecular Clouds as an Indication of Ambipolar Diffusion}
\titlerunning{Lifetime and Fragmentation Scale}

\author{Indrani Das
        \orcid{0000-0002-7424-4193}
         \inst{2, 1}
          \and
          Shantanu Basu
          \orcid{0000-0003-0855-350X}
          \inst{2}
          \and
          Philippe Andr\'e
          \orcid{0000-0002-3413-2293}
          \inst{3}
          }

\institute{Department of Applied Mathematics, University of Western Ontario, London, Ontario N6A 5B7, Canada\\
              \email{idas2@uwo.ca}
         \and
         Department of Physics and Astronomy, University of Western Ontario, London, Ontario N6A 3K7, Canada\\
             \email{basu@uwo.ca}
             \and
             Laboratoire d’Astrophysique (AIM), CEA/DRF, CNRS, Universit\'e Paris-Saclay, Universit\'e Paris Diderot, Sorbonne Paris Cit\'e,
91191 Gif-sur-Yvette, France\\
\email{philippe.andre@cea.fr}
             }
\date{}



\abstract{
Ambipolar diffusion likely plays a pivotal role in the formation and evolution of dense cores in weakly-ionized molecular clouds. Linear analyses show that the evolutionary times and fragmentation scales are significantly greater than the hydrodynamic (Jeans) values even for clouds with mildly supercritical mass-to-flux ratio.
We utilize values of fragmentation scales and growth times 
that correspond to typical ionization fractions 
within a molecular cloud, and 
apply to the context of the observed estimated lifetime of prestellar cores as well as the observed number of such embedded cores forming in a parent clump. By varying a single parameter, the mass-to-flux ratio, over the range of observationally measured densities, we fit the range of estimated prestellar core lifetimes ($\sim 0.1$ to a few Myr) identified with \textit{Herschel} as well as the number of embedded cores formed in a parent clump measured in Perseus with the \textit{Submillimeter Array} (SMA). Our model suggests that the prestellar cores are formed with a transcritical mass-to-flux ratio and higher densities correspond to somewhat higher mass-to-flux ratio but the normalized mass-to-flux ratio $\mu$ remains in the range $1 \lesssim \mu \lesssim 2$.
Our best-fit model exhibits $B \propto n^{0.43}$ for prestellar cores, due to partial flux-freezing as a consequence of ambipolar diffusion.

}


\maketitle 


\section{Introduction}

The condensation of dense structures
out of the diffuse interstellar medium (ISM) still has many mysteries. 
There are reasons to think that the influence of the magnetic field is preponderant. In recent years remarkable observational data has been obtained by \cite{planck2016} that allows a quantitative analysis of the relative orientation of the magnetic field within a set of nearby ($d < 450$ pc) molecular clouds. These observations have helped to establish the significance of magnetic-fields in the formation of density structures on physical scales ranging approximately $1-10 \, \rm{pc}$. They show a clear correlation of the direction of elongation of high density regions (number column density $N_{\rm H} \gtrsim 10^{22}$ cm$^{-2}$) becoming perpendicular to the ambient magnetic field direction.

Molecular clouds are known to contain hierarchichal nested density structures with, e.g., clumps, filaments, and cores \citep[see][]{Pandre2014, dobbs2014,heyer2015}.
\cite{pokhrel2018} did a study of hierarchical structure over five different scales (ranging from $\gtrsim \ $10$ \ {\rm pc} \ {\rm to} \ $10 \ {\rm AU}) in the Perseus molecular cloud using new observations from the \textit{Submillimeter Array} (SMA). 
They compared the number of fragments with the number of Jeans masses in each scale to calculate the Jeans efficiency, which is the ratio of observed to expected number of fragments.
\cite{konyves2015} used the results of the \textit{Herschel} Gould Belt survey 
\citep[HGBS-][]{andre2010} in the Aquila molecular cloud complex. 
They compared the numbers of 
prestellar cores in various density bins to the number of young stellar objects (YSOs), and estimated that the lifetime of prestellar cores is $\sim 1$ Myr, which is  $\sim 4$ times longer than the core free-fall time, and that the lifetime decreases as the average core density increases.
While current observations cannot determine whether ambipolar diffusion (neutral-ion slip) is occurring during the initiation of gravitational collapse,
nonideal magnetohydrodynamic (MHD) simulations suggest it plays an important role in establishing mildly supercritical mass-to-flux ratios in prestellar cores, whether starting from small-amplitude perturbations \citep[see][]{KudohBasu3D2007, basu09b} or
large-scale turbulent or converging flows \citep[see][]{nakamurali2005, basu09a, ChenOstriker2014}. 
Furthermore, in mildly supercritical regions, the hybrid modes driven by gravity and neutral-ion slip result in preferred lengthscales and growth times that can significantly exceed the Jeans scale and free-fall time, respectively \citep[see][]{basu04, ciolek06, bailey12}. 


In this Letter, we probe the variation of lifetime and fragmentation scales of dense structures in molecular clouds as a consequence of ambipolar diffusion. 
We apply the results of a linear analysis of ambipolar diffusion driven fragmentation in planar, isothermal, weakly-ionized, self-gravitating sheetlike magnetic clouds  
The calculated shortest growth timescale and preferred fragmentation mass for collapse are used to explain the observationally estimated lifetime of prestellar cores and the number of enclosed cores in a parent clump. 
In \autoref{sec:model} we describe the analytic model, in \autoref{sec:obs_fit} we present a comparison with the observational findings, and finally, in \autoref{sec:con} we summarize and draw conclusions from our results.

\section{Analytic Model}\label{sec:model}

We model interstellar molecular clouds as self-gravitating, partially ionized, isothermal, magnetic, planar, thin sheets with infinite extent in the $x$- and $y$-directions and a local vertical half-thickness $Z(x,y,t)$. Although sheets are an idealized geometry, a structured background state like a sheet or filament does capture the essential feature of a preferred scale of instability that is related to the local density scale length; this would not appear if assuming a uniform background. 
Note that the critical length scale and timescale can differ typically by a factor $\sim 2$ as compared to using a spherical (uniform) geometry. 
Our static initial state also does not include the effect of any large-scale motion that modifies the evolutionary timescale.
The nonaxisymmetric equations and formulations of the model have been described in detail in several papers \citep{ciolek06, bailey12}. 
The evolution equations include the nonideal MHD effect of ambipolar diffusion, the process by which neutrals are partially coupled to magnetic field through collisions with ions.
This effect is quantified by the neutral-ion collision (momentum-exchange) timescale \citep[e.g.,][]{BM94}:
\begin{equation}
\tau_{ni} \equiv 1.4 \frac{m_i + m_n}{m_i} \frac{1}{n_i \langle \sigma w \rangle_{i {\rm H}_2}},
\end{equation}
where ${\langle \sigma w \rangle}_{i \rm{H}_2}$ is the average collision rate between ions of mass $m_i$ (singly ionized Na, Mg, and HCO, adopted to have a
mass of 25 amu) and neutrals of mass $m_n \> (= 2.33 \> \rm{amu}$). We adopt a neutral-ion collision rate between ${\rm H}_2$ and $\rm{HCO}^{+}$ of $1.69 \times 10^{-9} \rm{cm}^3 \rm{s}^{-1}$ \citep{mcdaniel1973}. For the ion number density $n_i$ there is an assumed constant power-law approximation of the form $n_i \propto n_{n,0} ^{1/2}$ \citep{ciolek06, CM98powerk};
where $n_{n,0}$ is the initial uniform number density of neutrals.
The typical observed ionization fraction in molecular clouds (primarily due to cosmic ray ionization) \citep{elmegreen1979, Tielens2005book} is
\begin{equation}
    \chi_i =  10^{-7}\, \Bigg(\frac{n_{n,0}}{10^4 \; \rm{cm^{-3}}}\Bigg)^{-1/2} \, .
\label{eq:ion_frac}
\end{equation}
Such a low ionization means that ambipolar diffusion is unavoidable in molecular clouds, but there is still enough coupling with the charged species for the neutrals to be affected by the magnetic field. This is because of the high-polarizability of the neutrals, particularly $\rm{H}_2$ molecules \citep{Osterbrock1961}.
The threshold for whether a region of a molecular cloud is magnetically dominated or gravitationally dominated is given by the normalized mass-to-flux ratio,
\begin{equation}
\mu_0 \equiv 2\pi G^{1/2} \frac{\sigma_{n,0}}{B_{\rm{ref}}} ,
\label{eq:mu}
\end{equation}
where $\sigma_{n,0}$ is the initial uniform column density of the sheet, $B_{\rm{ref}}$ is the magnetic field strength of the background reference state, 
$G$ is the universal gravitational constant, and $(2\pi G^{1/2})^{-1}$ is the critical mass-to-flux ratio for gravitational collapse \citep{nakano1978}. 
Regions with $\mu_0 < 1$ are defined as subcritical, regions with $\mu_0 > 1$ are defined to be supercritical, and regions with $\mu_0 \approx 1$ are transcritical.
For small amplitude perturbations, the
governing equations 
can be combined to yield the following dispersion relation:

 \begin{equation}
\begin{aligned}
     \left(\omega + i \> \theta\right)  \big[\omega^2 - C^2_{\rm{eff,0}} \> k^2 + & 2\pi G \sigma_{n,0} k \big] = \\
                                                                                  & \omega \> \left[2\pi G \sigma_{n,0} k \mu_0 ^{-2} + k^2 \> V^2 _{A,0} \right] \> ,
\end{aligned}     
\label{eq:DR}     
\end{equation}
where 
\begin{equation}
    \theta = \tau_{ni,0} \left(2\pi G \sigma_{n,0}  \mu_0 ^{-2} k + k^2 V_{A,0} ^2 \right) = \etaAD (k \> {Z_0}^{-1}+ k^2),
\label{eq:theta}    
\end{equation} 
$\omega$ is an angular frequency, and $k^2  \equiv k_z^2= k_x^2 +k_y^2$, where $k_x$, $k_y$, $k_z$ (or $k$) are the wavenumbers in the $x$-, $y$- and $z$- directions, respectively. 
To obtain the dispersion relation as shown in \autoref{eq:DR}, the linearized perturbed MHD equations are used (see \autoref{sec:app1}). 
Here, $C_{\rm{eff,0}}$ and $V_{A,0}$ are the local effective sound speed and the Alfv\'en speed, respectively. The term
$C_{\rm{eff,0}}$ includes the effects of a restoring force due to an external pressure $P_{\rm{ext}}$. 
The dimensionless external pressure $\Tilde{P}_{\rm{ext}}$ ($\equiv 2 P_{\rm{ext}}/ (\pi G \sigma^2 _{n,0} )$) and temperature ($T$) are kept fixed at $0.1$ and $10 \ {\rm K}$, respectively.

\begin{figure}
\resizebox{\hsize}{!}{\includegraphics{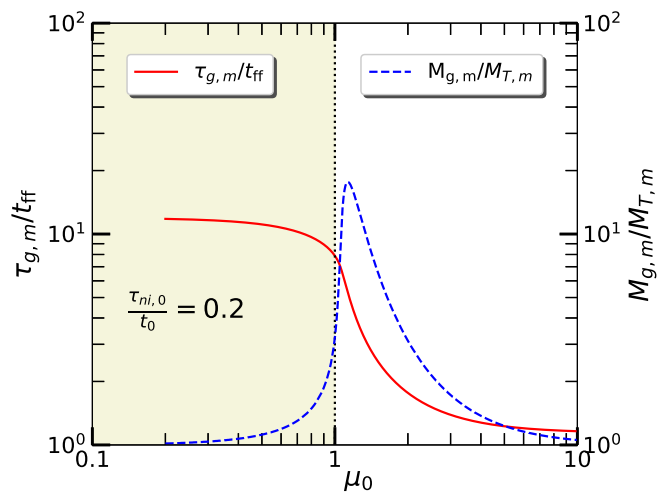}}
\caption{The growth timescale $\tau_{g,m}$ (in the units of dynamical or free-fall time $t_{\rm ff}$) and the preferred fragmentation mass $M_{g,m}$ (in the units of preferred thermal fragmentation mass $M_{T,m}$) of the most unstable mode as a function of the normalized mass-to-flux ratio ($\mu_0$). The model utilizes the standard ionization fraction corresponding to a normalized neutral-ion collision time $\tau_{ni,0}/t_0 = \taunit = 0.2$. The shaded and unshaded zones represent the subcritical ($\mu_0<1$) and supercritical ($\mu_0>1$) regimes, respectively. The present paper focuses on the (mildly) supercritical regime.}
\label{fig:tauni0p2}
\end{figure}

\autoref{fig:tauni0p2} presents the normalized shortest growth timescale, $\tau_{g,m}/t_{\rm ff}$ and normalized preferred fragmentation mass, $M_{g,m}/M_{T,m}$ corresponding to this minimum timescale as a function of $\mu_0$ for the case of normalized neutral-ion collision time $\tauni/t_0 = \taunit=0.2$ that corresponds to \autoref{eq:ion_frac}. 
Here, $t_{\rm ff}$ is the dynamical, i.e., free-fall time $(=Z_0/c_s)$, and $M_{T,m}$ is the preferred thermal mass based on our model and $M_{T,m} = (4\pi C^2_{{\rm eff},0}/c_s^2)^2 M_0$, where $M_0$ is effectively the Jeans mass.
See \autoref{sec:app1} for definitions and typical values of the units of time ($t_0$), length ($L_0$), and mass ($M_0$), and other parameters.
For all objects shown in \autoref{tab:numcore}, we calculate $M_{g,m}$ in units of $M_c = \pi \sigma_{n,0} (L_{0}/2)^{2} = \pi M_0/4$, as the perturbation is taken to be circular with radius $L_0/2$.
As the mass-to-flux ratio goes to the subcritical regime where ambipolar diffusion drives the evolution, the curve of shortest growth timescale approaches a plateau. 
It is noteworthy that the peak preferred fragmentation mass for collapse exceeds the Jeans mass by a factor of up to 10.  
Furthermore, for $\taunit = 0.2$ the timescale for collapse of a subcritical region is around $10 - 12$ times longer than that of a supercritical region; this is the origin of the often quoted result that the ambipolar diffusion time is $\approx 10$ times the free-fall time \citep{Mou1991, ciolek06, bailey12, Das_Basu_2021}. 
Hereafter, for a better representation we use $n_n$, $\sigma_n$, and $\mu$ instead of $n_{n,0}$, $\sigma_{n,0}$, $\mu_0$, respectively.
In this study, we are interested in the regime where the normalized mass-to-flux ratio $\mu$ remains in the range $1 \lesssim \mu \lesssim 2$. See \cite{KunzMou2009} for an application of the linear theory in the subcritical regime to model core masses.

\section{Observational Correspondence to Prestellar Cores}\label{sec:obs_fit}
In this section we discuss the relevance of our theoretical results, focused on the mildly supercritical regime, to the observational findings.

\subsection{Lifetime of Prestellar Cores}\label{sec:lifetime}
\begin{figure}
\resizebox{\hsize}{!}{\includegraphics{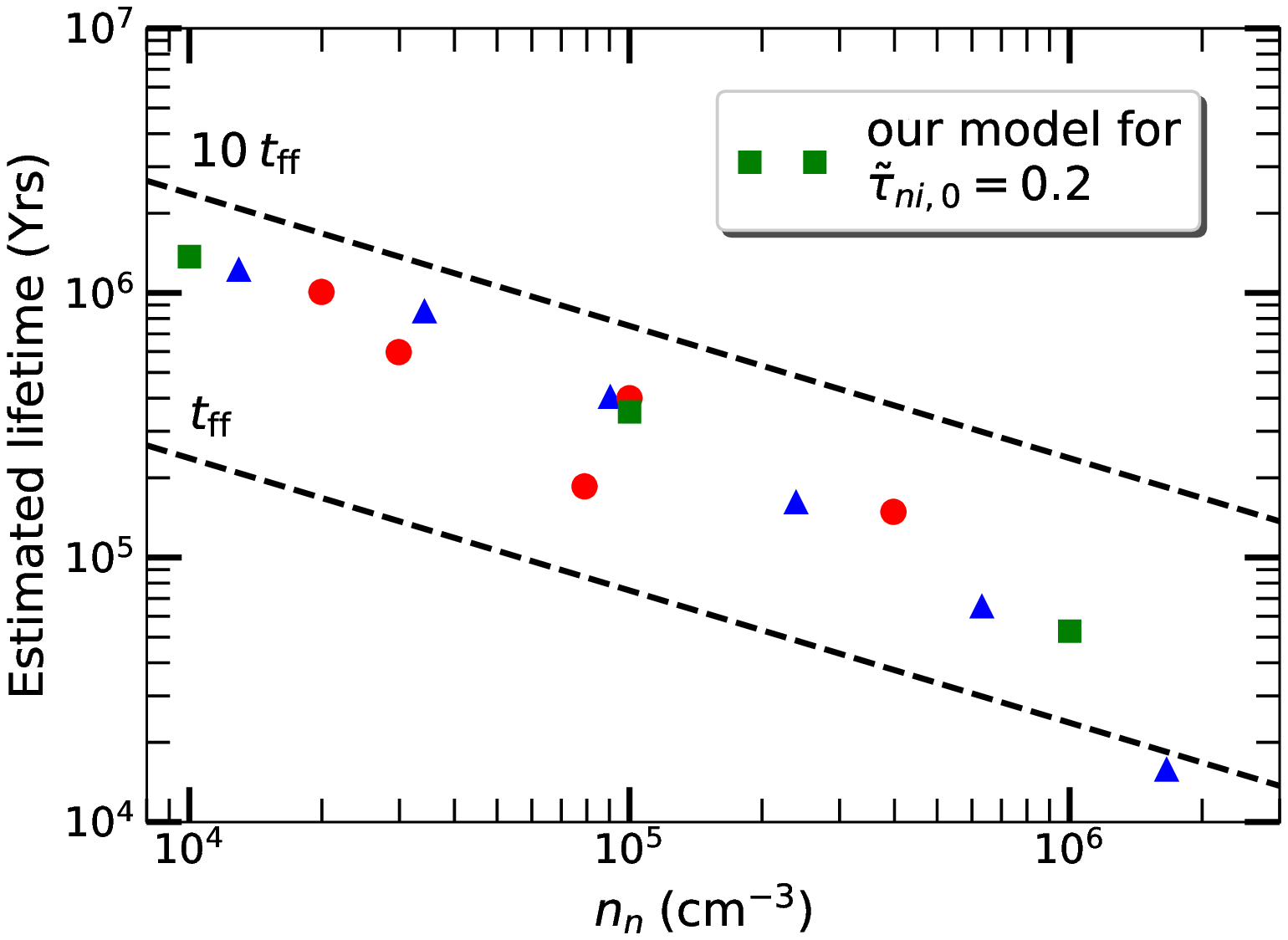}}
\caption{Estimated lifetime as a function of number density ($n_{n}$). The green filled squares are obtained from our model for a normalized neutral-ion collision time $\taunit=0.2$. The blue filled triangles show the corresponding data points for the population of 446 candidate prestellar cores identified with \textit{Herschel} in the Aquila cloud \citep{konyves2015}. Earlier data from \cite{WardThompson2007} is shown by the red circles for better comparison. The black dashed line shows the dynamical time ($Z_0/c_s$), i.e., the free-fall time, of our model.}
\label{fig:timescale_fit}
\end{figure}

The technique for finding the timescale of the core formation process was introduced by \cite{Beichman1986} in the context of IRAS sources, who studied the embedded YSOs within the core sample of \cite{MyersBenson1983} and \cite{myersetal1983}. They found that 35 cores had IRAS sources meeting the color selection criteria of embedded YSOs and 43 had no embedded IRAS sources. 
\cite{Beichman1986} calculated the percentage of cores with embedded sources to estimate the lifetime of a core without an embedded YSO by comparing it with the lifetime of the embedded YSO phase.
Using an estimated lifetime of cores with embedded class II sources of $1-2 \> {\rm Myr}$ \citep[as discussed in][]{WardThompson1994, Evans2009}, and assuming that the prestellar cores will go on to form protostars, one estimates a prestellar lifetime
\begin{equation}
    \tau = \frac{\rm \# \ of \ cores \ without \   embedded \ sources}{\rm \# \ of \ cores \ with \ embedded \ sources} \times \> {[1-2] \> {\rm Myr}}.
\end{equation}
This formula was used by \cite{JessopWardThompson2000} on a catalog of molecular cloud cores from the all-sky IRAS Sky Survey Atlas (ISSA), and their fig. 6 (often called a ``JWT'' plot) shows the estimated lifetime versus mean density.
In a similar way, \cite{konyves2015} estimated the lifetime of prestellar cores (see their fig. 9) by comparing the number of 
prestellar cores found with \textit{Herschel} to the number of Class II YSOs detected by \textit{Spitzer} in the Aquila cloud.
This study had the advantage of being a homogeneous sample of cores from a single cloud (Aquila), measured using a single telescope, tracer (dust), and analysis technique that separated prestellar (gravitationally bound) cores from starless (unbound) cores. Some of the underlying assumptions in these studies are: 1) that prestellar cores will evolve into YSOs in the future; 2) that star formation proceeds at a roughly constant rate, at least when averaged over an entire cloud. 

\autoref{fig:timescale_fit} presents the core lifetime estimated from our model as the instability growth time at a particular density $n_{n}$, when adopting a 
specific model for $B_{\rm ref}$ as a function of neutral number density (see \autoref{fig:B_fit} and \autoref{eq:B_n_fit} below). The green filled squares are our model values. The red filled circles represent the literature data from \cite{WardThompson2007}, and the blue filled triangles are the numbers estimated from a 
population of 446 candidate prestellar cores identified with \textit{Herschel} in the Aquila Cloud \citep{konyves2015}. 
We achieve the good correspondence by varying only one relationship, that between the normalized mass-to-flux-ratio $\mu$ and the number density $n_{n}$, with $\mu$ ranging from 
about 1.1 to 1.5 for $n_{n}$ between $10^4 \ - 10^6 \ {\rm cm}^{-3}$. To calculate $\mu$ we evaluate $\sigma_{n}$ using \autoref{eq:pbal}. These values of $\mu$ are in the range of mildly supercritical values generally obtained from Zeeman detections and use of the Davis-Chandrasekhar-Fermi (DCF) method \citep{crutcher2012,pattle2019}. The ionization level is set by the value $\taunit =0.2$ corresponding to the standard value set by \autoref{eq:ion_frac}.
Our use of the instability growth time as a proxy for the evolutionary time is similar to the commonly used comparison of the free-fall time at a particular density with the evolutionary time of a core at that density.
\autoref{fig:timescale_fit} shows that the typical lifetime of prestellar cores decreases from $\sim 1.37$ Myr for cores with volume density $\sim 10^4 \> \rm{ cm}^{-3}$ to $\sim 0.35$ Myr at $\sim 10^5 \> \rm{ cm}^{-3}$ to $\sim 0.05$ Myr at $\sim 10^6 \> \rm{ cm}^{-3}$ (see \autoref{tab:timescale}).
The timescale for collapse of a core of density $\sim 10^4 \> \rm{cm}^{-3}$ is around $10 - 50$ times longer than that of a highly dense core of $\sim 10^5-10^6 \> \rm{cm}^{-3}$. 
The highest density cores are significantly supercritical and evolve essentially on a free-fall timescale ($t_{\rm ff}=Z_0/c_s$) as shown in \autoref{fig:tauni0p2}. The lower black dashed line presents the dynamical timescale, or free-fall time, as a reference line. 



\subsection{Number of Enclosed Cores}\label{sec:numcore}

\begin{figure}
\resizebox{\hsize}{!}{\includegraphics{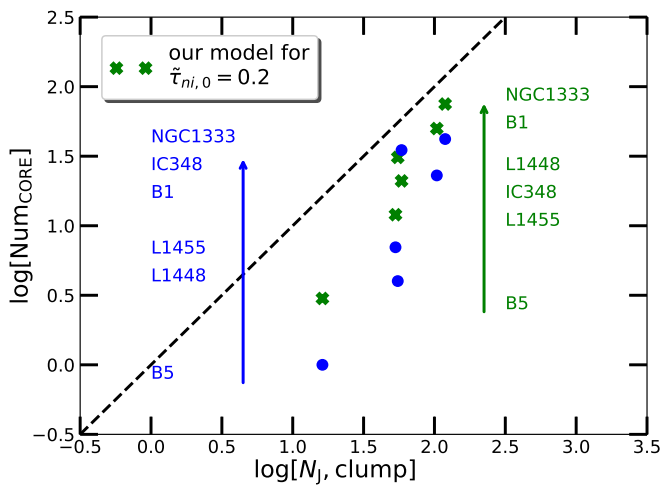}}
\caption{The number of enclosed cores ($\log[\rm{Num}_{\rm{CORE}}]$) as a function of Jeans number ($\log[N_{\rm{J, \> CLUMP}}]$) of clumps. The blue filled circles represent the observed number of enclosed cores as found by \cite{pokhrel2018}. The green filled crosses are obtained from our model for a normalized neutral-ion collision time $\taunit=0.2$. The black dashed line corresponds to an efficiency of unity.}
\label{fig:numcore_fit}
\end{figure}

\cite{Sadavoy2010} used point-source photometry 
to explore the dense cores in the Perseus star-forming complex as found in surveys with SCUBA $(85 \ \mu {\rm m})$ and \textit{Spitzer Space Telescope} $(3.6 - 70 \ \mu {\rm m})$.
\cite{mercimek2017} characterized the distribution of these cores inside the clumps. 
\cite{pokhrel2018} analyzed the submillimeter starless or protostellar cores in the \textit{Herschel} column density maps of \cite{mercimek2017} and used the estimated mass and areas (see \autoref{tab:numcore}) to determine the average density of each clump for $A_V > 7$ mag. They used the dust temperatures from \cite{Sadavoy2014} to estimate the thermal support.
To calculate the Jeans number ($N_{\rm{J, \> CLUMP}} \equiv M/M_{\rm J,th}$, i.e., the number of contained thermal Jeans masses $M_{\rm J,th}$) of the clumps, \cite{pokhrel2018} used the line-of-sight averaged temperatures and mass derived in \cite{Sadavoy2014}.

We fit the observations of the number of enclosed cores ($\rm{Num}_{\rm{CORE}}$) in each clump as a function of the corresponding Jeans number of the clumps,  as seen in \autoref{fig:numcore_fit}. 
We calculate $\rm{Num}_{\rm{CORE}}$ in the context of our model by dividing the total clump mass $M$ by the preferred fragmentation mass $M_{g,m}$, adopting $\taunit=0.2$. Note that $M_{g,m}$ significantly exceeds $M_{\rm J,th}$ for mildly supercritical objects. The Jeans number $N_{\rm{J, \> CLUMP}}$ is the expected number of cores in the context of thermally regulated fragmentation.
See \autoref{tab:numcore} for detailed specifications of all the clumps. 
The clumps are arranged in an increasing order of $\rm{Num}_{\rm{CORE}}$ (i.e., number of cores). 
 \autoref{fig:numcore_fit} shows that the number of cores (blue filled circles) increases with the Jeans number of the clumps, as shown in Figure 6 of \cite{pokhrel2018}. The filled green crosses represent our model values. The black dashed line shows the efficiency of unity ($\epsilon^{\rm{th}} =1$, i.e., $\rm{Num}_{\rm{CORE}}=N_{\rm{J, \> CLUMP}}$) corresponding to purely thermally regulated fragmentation. 

The observations show fewer cores than that predicted with only thermal pressure.
This hints at a relatively larger threshold for fragmentation mass that includes effects beyond that set by thermal pressure alone.
In our model, $M_{g,m}$ serves such a purpose as a magnetic field dependent instability threshold in contrast to a Jeans mass. 

It is worth noting that the number of fragmented cores is also dependent on the clump mass. 
The relatively massive clumps are able to generate more cores and $\rm{Num}_{\rm{CORE}}$ comes closer to (but stays below) the value of $N_{\rm{J, \> CLUMP}}$.
Therefore, the number density might not be the only key parameter in this context. The nonthermal motions in these massive clouds could also play a role. However, using the nonthermal dispersion in the calculation of the Jeans mass would not fit the observations for these clouds either, predicting many fewer cores than are observed \citep[see the discussion in][]{pokhrel2018}.

The inability of the observed nonthermal dispersion to be used as the source of an internal pressure when estimating the number of fragmented cores could possibly be attributed to the following reasons. First, the nonthermal motions may arise at least in part due to large-scale velocity gradients, in which case they cannot act as an effective pressure. Simulations in turbulent boxes that have global stability show that small scale collapse still occurs unless the driving scale and/or power spectrum is peaked at extremely short scales \citep[e.g.][]{vaz96,kle00}. Second, the nonthermal dispersion may be dominated by motions in the lowest densities that are traced, while a dense layer that actually undergoes fragmentation may have a lower more nearly thermal dispersion. The latter explanation is supported by simulations \citep{kud03,kud06,fol04} that show that the velocity dispersion peaks in the low density regions of a stratified molecular cloud. In observations as well, the velocity dispersion peaks in low-density regions and starless dense cores correspond to minima in velocity dispersion maps \citep[see, e.g., fig. 13 of][]{FriesenPineda2017}.


\subsection{Estimation of Magnetic Field}

\begin{figure}
\resizebox{\hsize}{!}{\includegraphics{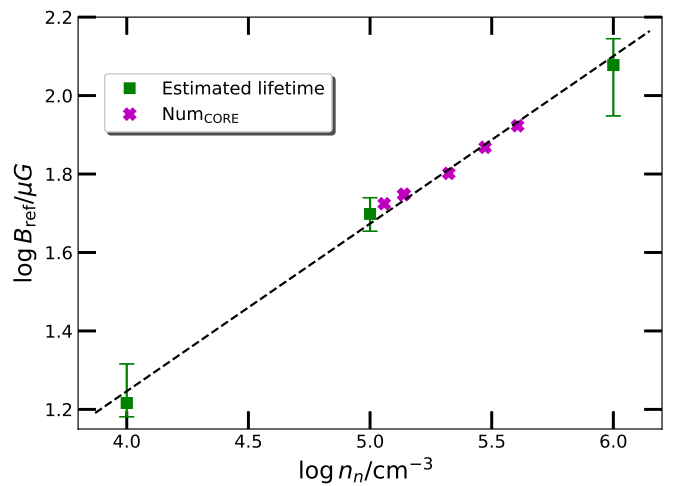}}
\caption{Magnetic field $\log (B_{\rm ref}/\mu \rm{G})$ versus number density $\log (n_{n}/{\rm cm}^{-3})$. The filled squares and crosses represent the number density region corresponding to the fitting of the lifetime of prestellar cores and $\rm{Num}_{\rm{CORE}}$ as shown in \autoref{fig:timescale_fit} and \autoref{fig:numcore_fit} (see \autoref{tab:timescale} \& \autoref{tab:numcore}). The dashed line is the least-squares fit.
The vertical error bars correspond to magnetic field variations that yield a total factor of 2 (greater or lesser by a factor $\sqrt{2}$) values of the growth time (lifetime).}
\label{fig:B_fit}
\end{figure}

For gravitationally contracting fragments (particularly cores) in magnetic interstellar clouds, the relation between magnetic-field strength $B_{\rm ref}$ and gas number density $n_{n}$ is of considerable interest.
In \autoref{fig:B_fit}, we present $\log (B_{\rm ref}/\mu G)$ (which is obtained from our model) as a function of $\log (n_{n}/{\rm cm}^{-3})$ for the two different density regimes shown in \autoref{fig:timescale_fit} and \autoref{fig:numcore_fit} based on our model for $\taunit =0.2$.
We use the least-squares method to find a best fit to the data:
\begin{equation}
    \frac{B_{\rm ref}}{10 \ \mu {\rm G}} = A \left(\frac{n_{n}}{10^4\> {\rm cm}^{-3}}\right)^{\kappa}\, ,
\label{eq:B_n_fit}    
\end{equation}
where $\kappa = 0.427$ and $A = 1.766$, and it is shown by the black dashed line in \autoref{fig:B_fit}.

\cite{mestel1965} argued that for a spherically and isotropically collapsing magnetic cloud, the scaling between the magnetic field strength and the density is $B \propto n_{n}^{2/3}$. This argument is true for an quasi-spherical collapse where both the mass $M$ and the magnetic-flux $\Phi$ are being conserved, and the magnetic field energy is insignificant compared to the gravitational energy. Later, \cite{Mouschovias1976a,mouschovias1976b} argued that the plasma $\beta$ ($=  8\pi \rho c_s^2/B^2$) remains constant during self-contraction of a cloud with a dynamically important magnetic field, therefore $B \propto n_{n}^{1/2}$, although for individual points within the cloud the exponent was in the range $1/3-1/2$. These theories assume an evolutionary sequence and that the mass-to-flux ratio and internal thermal (or turbulent) velocity dispersion stays fixed. 
We should keep in mind that the full virial relation is 
\begin{equation}
    B \propto \sigma\, n_{n}^{1/2} / \mu \>,
\end{equation}
where $\sigma$ is the velocity dispersion. With
the inclusion of ambipolar diffusion and a systematically increasing value of $\mu$ as $n_{n}$ increases, the slope in the $B_{\rm ref}-n_n$ relation is expected to be less than in the flux-freezing models. On the other hand, a systematic dependence of $\sigma$ on $n_{n}$ can also have an effect. \cite{crutcher2010} fitted a slope of $\approx 2/3$ to Zeeman magnetic field data for an ensemble of dense molecular gas clouds. 
The $B-n$ relation shown in \autoref{fig:B_fit} is for prestellar cores while the $B-n$ relation shown by \cite{crutcher2010} includes a number of massive protostellar cores or clumps.
Those objects definitely do not represent an evolutionary sequence, with the higher density objects representing much more massive clouds that also have an increased velocity dispersion $\sigma$. A better fit to the ensemble of different clouds measured by the Zeeman effect is $B \propto \sigma n^{1/2}$ \citep{basu2000} and was verified by \cite{li2015} who used the updated data in \cite{crutcher2012} and pointed out that the ensemble of objects in the data set also have the correlation $\sigma \propto n^{1/6}$, thereby leading to an apparent $B \propto n^{2/3}$. A similar result is obtained for $B$ determined through the DCF technique \citep{MyersBasu2021}.

\section{Conclusions}\label{sec:con}

We have utilized a semi-analytic model of ambipolar-diffusion driven
gravitational fragmentation in isothermal self-gravitating interstellar molecular clouds.
The only requirements in our model are that prestellar cores are \textit{transcritical} (mildly supercritical), with $1 \lesssim \mu \lesssim 2$ and an evolution toward the higher values of $\mu$ as the density increases.
With this assumption we have shown that there is a significant and systematic variation of lifetime and fragmentation scale in molecular clouds. 
Such systematic variations do not exist in the standard thermal pressure dominated (Jeans) fragmentation theory and are difficult to reproduce in turbulence-regulated models.
Our best-fit model for prestellar cores suggests $B \propto n^{0.43}$ (see \autoref{fig:B_fit} and \autoref{eq:B_n_fit}), which attains a shallower slope than the flux-frozen case due to the effects of ambipolar diffusion.
The estimated lifetime of prestellar cores and the possible number of cores within a parent clump/cloud based on the model for $\taunit =0.2$ agree well with that of observations presented by \cite{konyves2015} and \cite{pokhrel2018}, respectively.
For the lower end ($\sim 10^4 \> \rm{cm^{-3}}$) of the density regime, the timescale for collapse of prestellar cores is $\sim 6$ times
longer than the free-fall timescale. Whereas for the case of relatively higher number density ($\sim 10^6 \> \rm{cm^{-3}}$), the timescale is nearly as same as the free-fall timescale.
The mass scales of fragment formation are also significantly greater than the Jeans mass in this mildly supercritical regime.

We have adopted the dimensionless neutral-ion collision time $\taunit=0.2$ because of its correspondence to the typical ionization fraction ($\sim 10^{-7}$) at a neutral number density ($\sim 10^4 \> \rm{cm^{-3}}$). 
In future studies, the role of varying $\tilde{\tau}_{ni,0}$ could be explored in order to compare with measured ionization fractions in cores \citep[see][]{PC2002}, and perhaps constrain the cosmic ray ionization rate (canonical value $\zeta_{\rm CR} = 10^{-17} \, {\rm s}^{-1}$). 

Our model provides a means to indirectly infer the effect of ambipolar diffusion on mildly supercritical dense regions (prestellar cores) of molecular clouds. The importance of ambipolar diffusion in dense supercritical molecular cloud gas has not been discussed widely and is independent of its possible effects in low density molecular cloud envelopes.

\begin{acknowledgements} 
We thank the anonymous referee for comments that improved the manuscript.
SB is supported by a Discovery Grant from NSERC. 
PhA acknowledges support from “Ile de France” regional funding (DIM-ACAV+ Program)
and from the French national programs of CNRS/INSU on stellar and ISM physics (PNPS and PCMI).
\end{acknowledgements}


\bibliography{myref}
\bibliographystyle{aa} 

\begin{appendix}
\section{System of Equations}\label{sec:app1}
We formulate model clouds as rotating, self-gravitating,
partially ionized, isothermal, magnetic, planar thin sheets with
infinite extent in the $x$- and $y$- directions and a local vertical half-thickness $Z(x, y, t)$ (see Figure 1 from \cite{Das_Basu_2021}).
In our model, we adopt a velocity unit of $c_s$, the isothermal sound speed, and a column density unit of $\sigma_{n,0}$, the initial uniform column density. The length unit is $L_0 = c_s ^2/(2\pi G \sigma_{n,0})$, leading to a time unit $t_0= c_s/(2\pi G \sigma_{n,0})$, where $G$ is the universal gravitational constant.  
The mass unit is $M_0 = c_s^4/(4\pi^2 G^2 \sigma_{n,0})$ and
the magnetic field strength unit is $B_0 = 2\pi G^{1/2}\sigma_{n,0}$. 
The free-fall time is
\begin{equation}
    t_{\rm ff} = Z_0/c_s = (\tilde{Z}_0 L_0)/c_s = \tilde{Z}_0 \, t_0 \ ,
\end{equation}
where
\begin{equation}
    Z_0 = \sigma_{n,0}/ (2\rho_{n,0})
\end{equation}
is the initial uniform local vertical half-thickness, with dimensionless form
\begin{equation}
    \tilde{Z}_0 = \frac{2}{(1+ \tilde{P}_{\rm ext})} \ .
\end{equation}
Here $\tilde{P}_{\rm ext} = 2P_{\rm ext}/( \pi G \sigma^2_{n,0})$ is the dimensionless external pressure.
The initial local effective sound speed, $C_{\rm{eff,0}}$, comes from
\begin{equation}
    C^2_{\rm{eff,0}} = \frac{\pi}{2} G \sigma^2_{n,0} \frac{\left[3P_{\rm{ext}}+ \frac{\pi}{2} G \sigma^2_{n,0}\right]}{\left[P_{\rm{ext}}+\frac{\pi}{2} G \sigma^2_{n,0}\right]^2} c_s ^2,
\end{equation}
and it reduces to the isothermal sound speed, $c_s$, in the limit of very low $P_{\rm{ext}}$. It can be obtained from the linearized condition of vertical hydrostatic equilibrium:
\beq
\rho_{n,0} c_s ^2 = \frac{\pi}{2}G \sigma_{n,0} ^2 + \> P_{\rm{ext}} = \frac{\pi}{2}G \sigma_{n,0} ^2 (1 + \tilde{P}_{\rm ext})\, .
\label{eq:pbal}
\eeq
The initial uniform Alfv\'en speed $V_{A,0}$ is related to the mass-to-magnetic-flux ratio $(\mu_0)$ via
\begin{equation}
    V^2_{A,0} \equiv \frac{B^2_{\rm{ref}}}{4 \pi \rho_{n,0}} = 2\pi G \sigma_{n,0} \mu_0 ^{-2} Z_0 \>\>,
\end{equation}
where $\rho_{n,0}$ is the initial uniform volume density. 
The initial uniform ambipolar diffusivity can be expressed as \citep[see][]{Das_Basu_2021}
\begin{equation}
    \etaAD = V^2_{A,0} \tau_{ni,0} = 2\pi G \sigma_{n,0} \mu_0 ^{-2} Z_0  \tauni.
\end{equation}
Now, from the system of governing equations (see Equations (11) to (23) of \cite{Das_Basu_2021}, which are the fundamental MHD equations obtained by integrating over the $z$-direction),
we obtain a simplified form in $x$- and $y$- components:
\begin{equation}
     \frac{\partial \sigma_n}{\partial t} + \frac{\partial}{\partial x}(\sigma_n v_{n,x}) + \frac{\partial}{\partial y} (\sigma_n v_{n,y}) = 0 ,
\label{eq:masscontequn_xy}
\end{equation}

\begin{equation}
\begin{aligned}
    \frac{\partial}{\partial t} (\sigma_n v_{n,x}) {} & + \frac{\partial}{\partial x} (\sigma_n v_{n,x}^2)+ \frac{\partial}{\partial x}  (\sigma_n v_{n,x} v_{n,y})\\
                                                     &  = \sigma_n \> g_x - C_{\rm{eff}} ^2 \frac{\partial \sigma_n}{\partial x} + F_{{\rm Mag}, x} + 2 \sigma_n \Omega v_{n,y},
\end{aligned}   
\label{eq:forceequn_x}
\end{equation}

\begin{equation}
\begin{aligned}
    \frac{\partial}{\partial t} (\sigma_n v_{n,y}) {} & + \frac{\partial}{\partial y} (\sigma_n  v_{n,x} v_{n,y})+ \frac{\partial}{\partial y}  (\sigma_n v_{n,y}^2) \\
                                                     & = \sigma_n \> g_y - C_{\rm{eff}} ^2 \frac{\partial \sigma_n}{\partial y} + F_{{\rm Mag}, y} - 2 \sigma_n \Omega v_{n,x},
\end{aligned}
\label{eq:forceequn_y}
\end{equation}

\begin{equation}
\begin{aligned}
    F_{{\rm Mag},x} {} & = \frac{B_{z, \rm{eq}}}{2\pi} \> \left(B_{x,Z} - Z\> \frac{\partial B_{z, \rm{eq}}}{\partial x} \right) \\
                      & + \frac{1}{4\pi} \frac{\partial Z}{\partial x} \Biggr[B^2_{x,Z} +B^2_{y,Z} +  2 B_{z,\rm{eq}} \left(B_{x,Z} \frac{\partial Z}{\partial x}+ B_{y,Z} \frac{\partial Z}{\partial y} \right)\\
                      & \hspace{3 cm}+ \left(B_{x,Z} \frac{\partial Z}{\partial x}+ B_{y,Z} \frac{\partial Z}{\partial y} \right) ^2 \Biggr],
\end{aligned}          
\label{eq:fmagx}
\end{equation}

\begin{equation}
\begin{aligned}
    F_{{\rm Mag},y} {} & = \frac{B_{z, \rm{eq}}}{2\pi} \> \left(B_{y,Z} - Z\> \frac{\partial B_{z, \rm{eq}}}{\partial y}\right) \\
                      & + \frac{1}{4\pi} \frac{\partial Z}{\partial y} \Bigg[B^2_{x,Z} +B^2_{y,Z} +  2 B_{z,\rm{eq}} \left(B_{x,Z} \frac{\partial Z}{\partial x}+ B_{y,Z} \frac{\partial Z}{\partial y}\right)\\
                      & \hspace{3 cm} + \left(B_{x,Z} \frac{\partial Z}{\partial x}+ B_{y,Z} \frac{\partial Z}{\partial y} \right)^2 \Bigg],
\end{aligned} 
\label{eq:fmagy}
\end{equation}

\begin{equation}
    \begin{aligned}
    \frac{\partial B_{z,\rm{eq}}}{\partial t} = -\frac{\partial}{\partial x} {} & \left(B_{z,\rm{eq}} v_{i,x} \right)  -\frac{\partial}{\partial y} \left(B_{z,\rm{eq}} v_{i,y} \right)\\
                                                                                & + \left[\frac{\partial}{\partial x}\left(\eta_{\rm{OD}} \frac{\partial B_{z,\rm{eq}}}{\partial x}\right)+ \frac{\partial}{\partial y} \left(\eta_{\rm{OD}} \frac{\partial B_{z,\rm{eq}}}{\partial y}\right)\right].
    \end{aligned}
\label{eq:inductionequn_xy}    
\end{equation}
where
\begin{equation}
    v_{i,x} = v_{n,x} + \frac{\tau_{ni,0}}{\sigma_n} \> \left(\frac{\rho_{n,0}}{\rho_n} \right)^{1/2} \> {F_{{\rm Mag},x}} ,
\label{eq:velocityequn_in_xcomp} 
\end{equation}

\begin{equation}
    v_{i,y} = v_{n,y} + \frac{\tau_{ni,0}}{\sigma_n} \> \left(\frac{\rho_{n,0}}{\rho_n} \right)^{1/2} \> {F_{{\rm Mag},y}}.
\label{eq:velocityequn_in_ycomp}    
\end{equation}
Here, $v_{i,x}$, $v_{i,y}$, $v_{n,x}$, and $v_{n,y}$ are the $x$- and $y$- components of ion and neutral velocities. 
The planar sheet is rotating with an angular velocity $\Omega$ about the $z$-axis, so that $\vec{\Omega} = \Omega \hat{z}$. The magnetic field and rotation axis are perpendicular to the sheet. Here, $\etaOD$ is the Ohmic diffusivity that is considered as a measure of Ohmic dissipation.
Starting with a static uniform background, any physical quantity of the thin-sheet equations can be expanded by
writing it via 
\beq
f(x,y,t) = f_0 + \delta f_a e^{i\left( k_x x + k_y y -\omega t \right)},
\eeq
where $f_0$ is the unperturbed background state, $\delta f_a$ is the amplitude of the perturbation. 
$k_x$, $k_y$, and $k$ are the $x$-, $y$-, and $z$- wavenumbers, respectively, and $\omega$ is the complex angular frequency. 
For assumed small-amplitude perturbations such that $\left|\delta \>f_a \right| \ll f_0 $, and retaining the linearized form of the perturbed quantities from Eqs. \ref{eq:masscontequn_xy}, 
\ref{eq:forceequn_x}, \ref{eq:forceequn_y} and \ref{eq:inductionequn_xy}, the following equations are obtained: 
\beq
\omega \> \delta \sigma'_{n} =  k_x \> c_s \> \delta v'_{n,x} + k_y \>  c_s \> \delta v'_{n,y} \>\>,
\label{eq:lin_mass_con}
\eeq

\beq
\begin{aligned}
\omega \> c_s \> \delta v'_{n,x} ={} & \frac{k_x}{k} \> \left[C_{\rm{eff,0}} ^2 \> k - 2 \pi G \sigma_{n,0} \right] \> \delta \sigma'_{n} + i \> 2\Omega c_s \delta v'_{n,y} \\
                                    & + \frac{k_x}{k} \> \left[\> 2 \pi G \sigma_{n,0} \> \mu_0 ^{-1} + k\>  V_{A,0} ^2 \> \mu_0 \>\right] \>\delta B'_{z,\rm{eq}} \>\>,
\end{aligned}
\label{eq:lin_force_x}
\eeq
\beq
\begin{aligned}
\omega \> c_s \> \delta v'_{n,y} ={} &  \frac{k_y}{k} \> \left[C_{\rm{eff,0}} ^2 \> k - 2 \pi G \sigma_{n,0}\right] \> \delta \sigma'_{n} - i \> 2\Omega c_s \delta v'_{n,x} \\
                                    & + \frac{k_y}{k} \> \left[\> 2 \pi G \sigma_{n,0} \> \mu_0 ^{-1} + k\>  V_{A,0} ^2 \> \mu_0 \>\right] \>\delta B'_{z, \rm{eq}} \>\>,
\end{aligned}
\label{eq:lin_force_y}
\eeq

\beq
\begin{aligned}
\omega \> {} & \delta B'_{z, \rm{eq}}  = \frac{k_x}{\mu_0} \> c_s \> \delta v'_{n,x}  +  \frac{k_y}{\mu_0} \> c_s \> \delta v'_{n,y} \\
             & - i \left[\etaOD \> k^2 + \>\tau_{ni,0} \> \left(2\pi G \sigma_{n,0} \mu_0 ^{-2}k + k^2 \> V_{A,0}^2 \right)\right] \> \delta B'_{z,\rm{eq}} ,
\end{aligned}
\label{eq:lin_induction}
\eeq
where 
the perturbed eigenfunctions $\delta \sigma_{n}$, $\delta v_{n,x}$  (and $\delta v_{n,y}$), $\delta B_{z, \rm{eq}}$ are normalized by $\sigma_{n,0}$, $c_s$, and $B_0 \, (= 2\pi G^{1/2}\sigma_{n,0})$, respectively such that $\delta \sigma'_{n} = \delta \sigma_{n}/\sigma_{n,0}$, $\>\delta v'_{n,x} = \delta v_{n,x}/c_s$, $\delta v'_{n,y} = \delta v_{n,y}/c_s \> $, and $\> \delta B'_{z, \rm{eq}} = \delta B_{z, \rm{eq}}/B_0$. 
Now, finding the determinant from the above set of equations one obtains the full dispersion relation
\begin{equation}
\begin{aligned}
     \left(\omega + i \> [\theta + \gamma] \right)  \big(\omega^2 {} & - C^2_{\rm{eff,0}} \> k^2  + 2\pi G \sigma_{n,0} k - 4\Omega^2 \big) \\
                                                     & = \omega \> \left[2\pi G \sigma_{n,0} k \mu_0 ^{-2} + k^2 \> V^2 _{A,0} \right] \;.
\end{aligned}
\label{eq:fullDR}
\end{equation}
This is derived in \cite{Das_Basu_2021} (see Equation (47)), where 
\begin{equation}
    \gamma = \etaOD \> k^2 \ ,
\end{equation}
which is same as Equation (45) of \cite{Das_Basu_2021}, 
and $\theta$ is described earlier (see Equation \ref{eq:theta} and Equation (46) of \cite{Das_Basu_2021}).
Finally, the dispersion relation (see Equation \ref{eq:DR}) is the same as the Equation \ref{eq:fullDR} when setting $\Omega=0$ and $\etaOD=0$.
In the above, we discuss all the equations in detail for completeness and clarity of our model. 
See also Equations 32(a)–32(d) of \cite{ciolek06} for the dimensionless representation of the \Cref{eq:lin_mass_con,eq:lin_force_x,eq:lin_force_y,eq:lin_induction} for the model with $\Omega=0$ and $\etaOD=0$; also see Equations (10) to (13) of \cite{bailey12} for the dimensional representation of \Cref{eq:lin_mass_con,eq:lin_force_x,eq:lin_force_y,eq:lin_induction} for the same model.

Based on these parameters, typical values of the units used and other derived quantities are
\begin{equation}
    \sigma_{n,0}= \frac{3.63 \times 10^{-3}}{\left(1+\Tilde{P}_{\rm{ext}}\right)^\frac{1}{2}} \> \left(\frac{n_{n,0}}{10^3 \, \rm{cm}^{-3}}\right)^\frac{1}{2} \> \left(\frac{T}{10 \, \rm{K}}\right)^\frac{1}{2} \> \>  \rm{g} \> \rm{cm}^{-2},
\end{equation}

\begin{equation}
    L_0 = 7.48 \times 10^{-2}\> \left(\frac{T}{10 \, \rm{K}}\right)^ \frac{1}{2} \> \left(\frac{10^3 \, \rm{cm}^{-3}}{n_{n,0}}\right)^\frac{1}{2}\> \> \left(1+\Tilde{P}_{\rm{ext}}\right)^\frac{1}{2} \> \> \rm{pc},
\end{equation}

\begin{equation}
    t_0 = 3.98\>  \times 10^{5} \> \left(\frac{10^3 \, \rm{cm}^{-3}}{n_{n,0}}\right)^\frac{1}{2}\>\> \left(1+\Tilde{P}_{\rm{ext}}\right)^\frac{1}{2} \> \> \rm{yr},
\end{equation}

\begin{equation}
    c_s = 0.188 \> \left(\frac{T}{10\, \rm{K}}\right)^\frac{1}{2} \> \> \rm{km} \>\> \rm{s}^{-1}, 
\end{equation}

\begin{equation}
\begin{aligned}
    M_0 = 9.76 \times 10^{-2} \left(\frac{T}{10\, \rm{K}}\right)^{3/2} \; \left(\frac{10^3\,\rm{cm}^{-3}}{n_{n,0}}\right)^{1/2} \left(1+\Tilde{P}_{\rm{ext}}\right)^{1/2}  \> \> \> M_\odot\ ,
    \end{aligned}
\end{equation}

\begin{equation}
    \tau_{ni, 0}= \frac{3.74 \times 10^4}{\left(1+\Tilde{P}_{\rm{ext}}\right)} \left(\frac{T}{10\, \rm{K}}\right)  \left(\frac{0.01 \> \rm{g}\> \rm{cm}^{-2}}{\sigma_{n,0}}\right)^2 \; \left(\frac{10^{-7}}{\chi_{i,0}}\right) \> \> \rm{yr} \ ,
\end{equation}
 
\begin{equation}
\begin{aligned}
    B_{\rm{ref}} = \frac{5.89 \times 10^{-6}}{\mu_0} \left(\frac{n_{n,0}}{10^3\, \rm{cm}^{-3}}\right)^{1/2} \>\left(\frac{T}{10\, \rm{K}}\right)^{1/2} \left(1+\Tilde{P}_{\rm{ext}}\right)^{-1/2}\> \; {\rm G} \ .
\end{aligned}
\end{equation}


\begin{figure}
\resizebox{\hsize}{!}{\includegraphics{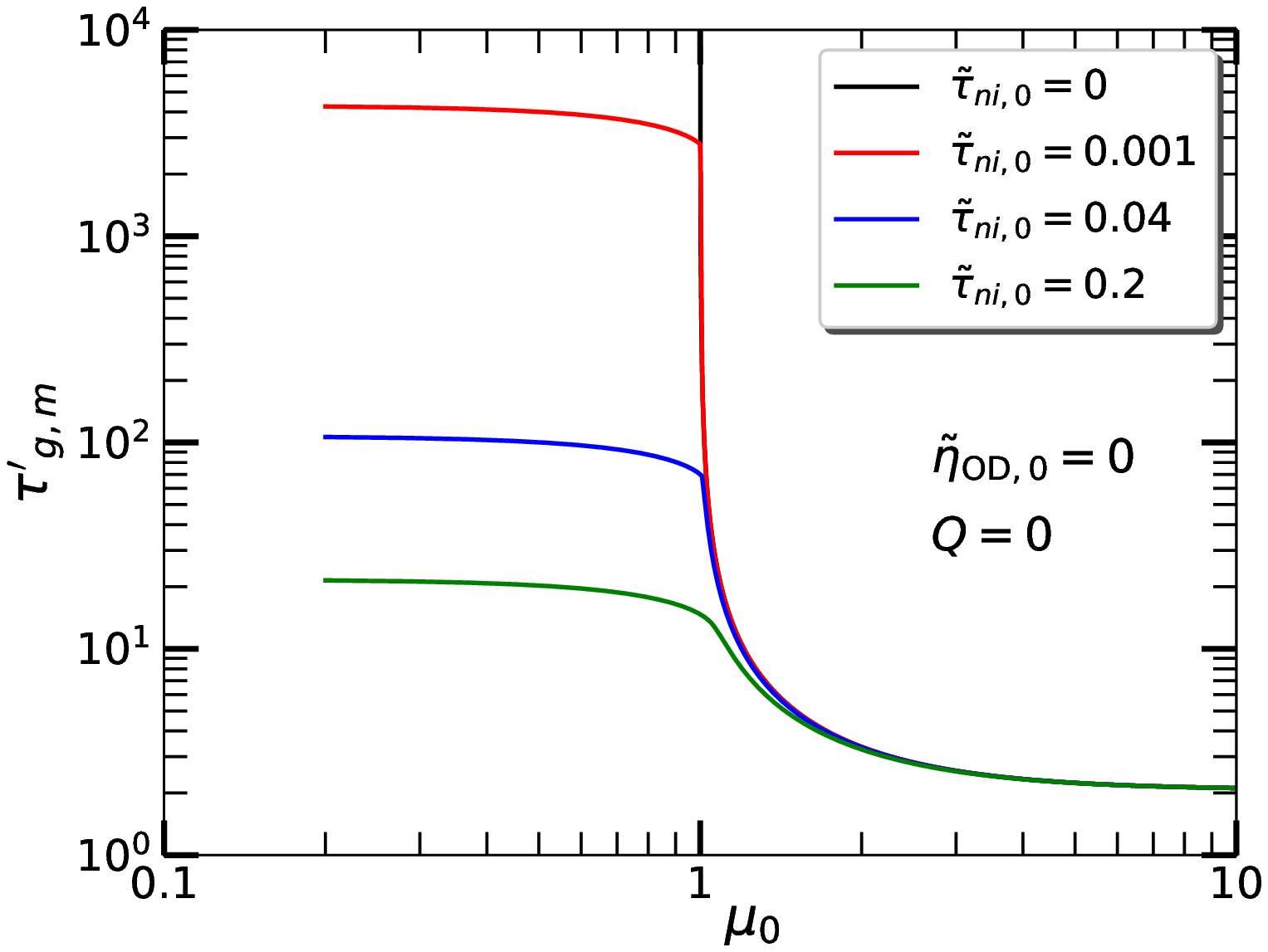}{(a)}}
\resizebox{\hsize}{!}{\includegraphics{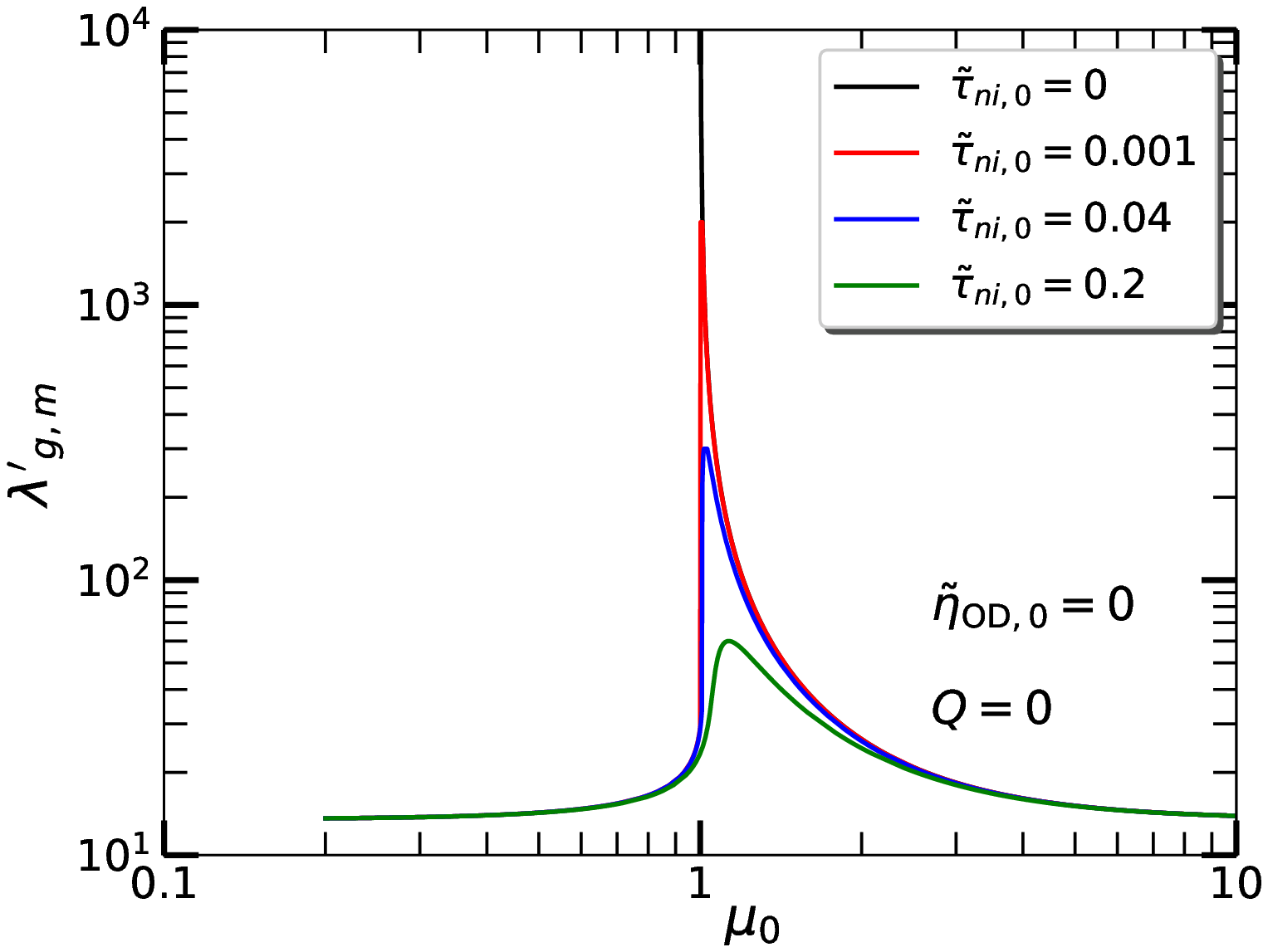}{(b)}}
\caption{Top: Normalized shortest growth time of gravitationally unstable mode ($\tau'_{g,m}=\tau_{g,m}/t_0$) as a function of normalized mass-to-flux ratio ($\mu_0 $). Bottom: Normalized preferred lengthscale of the most unstable mode ($\lambda'_{g,m} = \lambda_{g,m}/L_0$) as a function of the normalized mass-to-flux ratio ($\mu_0$). Each panel shows curves for models (as shown in \autoref{eq:DR}) with normalized neutral-ion collision time $\tilde{\tau}_{ni,0}$ = 0 (black), 0.001 (red), 0.04 (blue), and 0.2 (green). Here, $\tilde{\tau}_{ni,0}$ = 0 represents the flux-frozen case as a reference. The Toomre-$Q$ rotation parameter and the dimensionless Ohmic diffusivity $\etaODt$ are set to be zero. As the degree of ambipolar diffusion (i.e., $\taunit$) increases, the growth timescale and lengthscale become shorter \citep[see][]{ciolek06,bailey12,Das_Basu_2021} for further details.}
\label{figure:ADonly}
\end{figure}


\newpage
\section{Tables}\label{sec:app3}
\begin{table*}[htb!]
\caption{Fitting Data for Calculating the Lifetime of Prestellar Cores}
\label{table:timescale}
\centering 
\begin{tabular}{c c c c c c c c}
\hline\hline
$n_{n}$ & $\mu$ & $\sigma_{n}$ & $B_{\rm ref}$ ($\mu$G) & Estimated lifetime of\\ 
($\times\, 10^5\, \rm{cm}^{-3}$) & & ($\times \, 10^{-2}$ g  $\rm{cm}^{-2}$) & &  prestellar cores (Myr)\\
\hline 
0.1 & 1.080 & 1.094 & 16.45 & 1.371 \; ($\sim 5.777 \; t_{\rm ff}$)\\
1 & 1.125  & 3.461 & 49.94 & 0.355 \; ($\sim 4.728 \; t_{\rm ff}$)\\
10 & 1.486 & 10.944 &  119.57 & 0.053 \; ($\sim 2.220 \; t_{\rm ff}$)\\
\hline   
\hline
\end{tabular}
\label{tab:timescale}
\tablefoot{
\tablefoottext{a}{Here, $\mu$ is a free parameter of our model, $\sigma_{n}$ is calculated using \autoref{eq:pbal}, $B_{\rm ref}$ is obtained from \autoref{eq:mu}, and $t_{\rm ff}$ is the free-fall timescale. The estimated lifetime of prestellar cores is obtained based on \autoref{fig:tauni0p2}, as shown in \autoref{fig:timescale_fit}.} \\
\tablefoottext{b}{Table B.1 is arranged in the descending order of estimated lifetime of prestellar cores.}
}
\end{table*}




\begin{table*}[htb!]
\caption{Fitting Data for Calculating the Number of Enclosed Cores (${\rm Num}_{\rm CORE}$)}
\centering 
\begin{tabular}{c c c c c c c c c c c c c}
\hline\hline
Clump & Clump & Area & $\sigma_{n} \times 10^{-2}$ & $n_{n} \times 10^5$ & $\mu$ & $B_{\rm ref}$ & $M_{g,m}$ & $\rm{N}_{\rm{J}} ^{\rm{th}}$ & $\rm{N}_{\rm{J}} ^{\rm{th,nth}}$ & ${\rm Num}_{\rm CORE}$ & $\epsilon^{\rm{th}}$ & $\epsilon^{\rm{th,nth}}$\\ 

name & Mass ($\Msun$) & ($\rm{pc}^2$) & ($\rm{g\ cm}^{-2}$) &  $(\rm{cm^{-3}})$ & & ($\mu \rm{G}$) & ($\Msun$) &  & & & \\
\hline 

B5 & 62 & 0.32 & 4.064 & 1.383 & 1.178 & 56.01 & 20.165 & 16.2 & 1.5 & 3 & 0.185 & 2.000\\
L1455 & 251 & 1.3 & 4.050 & 1.373 &  1.176 & 55.90 & 20.339 & 53.1 & 4.3 & 12 & 0.226 & 2.791 \\
IC348 & 511 & 2.9 & 3.697 & 1.144 &  1.132 & 53.00 & 23.858 & 58.6 & 6.4 & 21 & 0.358 & 3.281 \\
L1448 & 159 & 0.48 & 6.948 & 4.042 &  1.470 & 76.73 & 5.108  & 55.1 & 4.6 & 31 & 0.562 & 6.739\\
B1 & 598 & 2.5 & 5.018 & 2.107 & 1.286 & 63.34 & 11.738 & 103.9 & 9.5 & 50 & 0.481 & 5.263\\
NGC1333 & 568 & 2.0 & 5.958 & 2.971 & 1.379 & 70.13 & 7.508 & 119 & 10.5 & 75 & 0.630 & 7.142\\
\hline 
\hline 
\end{tabular}
\label{tab:numcore}
\tablefoot{
\tablefoottext{a}{Clump name, Clump Mass, Area, $\rm{N}_{\rm{J}} ^{\rm{th}}$, $\rm{N}_{\rm{J}} ^{\rm{th, nth}}$ are taken from Table 2 of \cite{pokhrel2018}.}\\
\tablefoottext{b}{In the above table, $n_{n}$, $\sigma_{n}$, $\mu$, $B_{\rm ref}$, $M_{g,m}$, $\rm{Num}_{\rm{CORE}}$,  $\epsilon^{\rm{th}}$, $\epsilon^{\rm{th,nth}}$ are evaluated based on our model. Here, $\sigma_{n}$ is the clump mass per unit area, $n_{n}$ is obtained using \autoref{eq:pbal}, $\mu$ is a free parameter of our model, $B_{\rm ref}$ is calculated using \autoref{eq:mu}, and $M_{g,m}$ is obtained based on \autoref{fig:tauni0p2},
${\rm Num}_{\rm{CORE}} = {\rm Clump \ Mass}/M_{g,m}$, $\epsilon^{\rm{th}} = \rm{Num}_{\rm{CORE}}/\rm{N}_{\rm{J}} ^{\rm{th}}$, $\epsilon^{\rm{th,nth}} = \rm{Num}_{\rm{CORE}}/\rm{N}_{\rm{J}} ^{\rm{th,nth}}$}.\\
\tablefoottext{c}{Table B.2 is arranged in the ascending order of ${\rm Num}_{\rm{CORE}}$.}
}
\end{table*}

\end{appendix}

\end{document}